\documentstyle[aps,multicol,epsfig,floats]{revtex}

\begin{document}
\draft

\title{Thermopower of atomic-size metallic contacts}
\author{B. Ludoph and J.M. van Ruitenbeek}
\address{Kamerlingh Onnes Laboratorium, Universiteit Leiden, Postbus 
9504, NL-2300 RA Leiden, The Netherlands} 

\maketitle

\begin{abstract}
The thermopower and conductance of atomic-size metallic contacts have 
been simultaneously measured using a mechanically controllable break 
junction. For contacts approaching atomic dimensions, abrupt steps in the 
thermopower are observed which coincide with jumps in the conductance. 
The measured thermopower for a large number of atomic size contacts is 
randomly distributed around the value for large contacts and can be either 
positive or negative in sign. However, it is suppressed at the quantum 
value of the conductance $G_{0} = 2e^{2}/h$. We derive an expression that 
describes these results in terms of quantum interference of electrons 
backscattered in the banks. 
\end{abstract}



In recent years stable metallic contacts consisting of a single atom 
have become experimentally accessible \cite{1atoms}. The interesting interplay 
between quantization of the electron modes and the atomic structure of the 
contacts has resulted in intensive research in this field. Information obtained 
from experiments on atomic-size metallic contacts has mainly been limited to 
the measurement of the conductance.  Two exceptions stand out: the 
simultaneous measurements of force and conductance by Rubio {\it et al.}  \cite{rubio}, 
which prove that the conductance steps produced by contact elongation are due to 
atomic rearrangements, and the measurement of the subgap structure in atomic size 
superconducting aluminum contacts which characterize the conduction modes, by 
Scheer et al. \cite{scheer}. In this paper we present measurements of 
the thermopower in atomic-size metallic contacts. 

The thermopower $S$ is the constant of proportionality between an 
applied temperature difference $\Delta \theta$ and the induced voltage, $V_{tp} 
= S\Delta \theta$.  The relationship between the thermopower and the electrical 
conductance $G$ is given in the linear-response approximation by 
\[
S=-\frac{\pi ^{2}k_{B}^{2}\theta}{3e}\frac{\partial \ln G}{\partial \mu 
} ,
\]
with $\mu$ the chemical potential. One can view the thermopower as a measure for 
the difference in conductance between electron and hole quasiparticle excitations, 
or as the energy dependence of the conductance.  We will argue that the dominant 
contribution to the thermopower in atomic size contacts comes from quantum 
interference terms as a result of backscattering of electrons on defects near the contact. 
\begin{figure}[!b]
\begin{center}
\leavevmode
\epsfig{figure=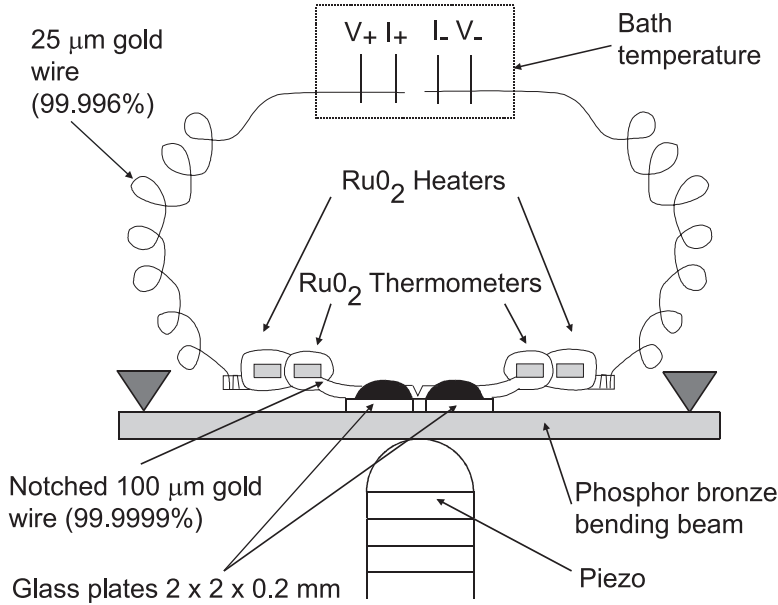,height=8cm,angle=0}
\end{center}
\caption{Schematic diagram of the modified MCB configuration, used for 
the simultaneous measurement of conductance and thermopower.}
\label{fig:fig1} 
\end{figure}
We have studied the thermopower of atomic-size gold contacts using a mechanically controllable break junction (MCB) \cite{muller}. A schematic diagram of the sample configuration is shown in Fig.\,\ref{fig:fig1}. 
By bending the phosphor bronze substrate, the 100\,$\mu$m gold wire breaks at the notch, allowing atomic-size contacts to be adjusted. This gold wire is attached to 
long thin (25\,$\mu$m) gold wires at both ends. They connect the notched wire to the current and voltage leads, anchored at the bath temperature, hence forming an open gold loop. The central gold wire is tightly wound and varnished on each side of the contact around a calibrated 5\,k$\Omega$ RuO$_{2}$ thermometer and a 500\,$\Omega$ RuO$_{2}$ heater. Using one heater, a temperature gradient over the contact can be applied. The glass plates and thin gold wires serve as thermal resistances to the substrate and bath temperatures, respectively. 
The sample is placed in a regular MCB setup in an evacuated can immersed in liquid helium. The conductance and thermopower were 
measured in three steps: The voltage over the contact was measured with a 
nanovoltmeter at $-$100\,nA, +100\,nA and at zero dc current bias, 
while maintaining a constant temperature gradient over the contact by 
applying about 2\,mW heating power to a heater on one side of the 
constriction. The conductance is then obtained from the voltage difference for the two current polarities, and $S$ is obtained from the voltage at zero 
bias current. Each cycle takes about 4\,s and is 
continuously repeated as we slowly sweep the piezovoltage up in order 
to decrease the contact size. A curve (Fig.\,\ref{fig:fig2}) was generally taken 
from 10\,$G_{0}$ to tunneling in 30 min ($G_{0}$ is the quantum conductance 
unit, $2e^{2}/h$).  Every few traces, it was necessary to readjust the contact 
manually, which inevitably leads to the contact being pushed completely together. Low pass RC filters (10\,Hz) were mounted in the circuit near the sample to prevent rectification of ac disturbances by the asymmetry of the voltage dependence of the conductance. Extensive measurements have been performed on two samples (referred to as samples 1 and 2). 
\begin{figure}[!t]
\begin{center}
\leavevmode
\epsfig{figure=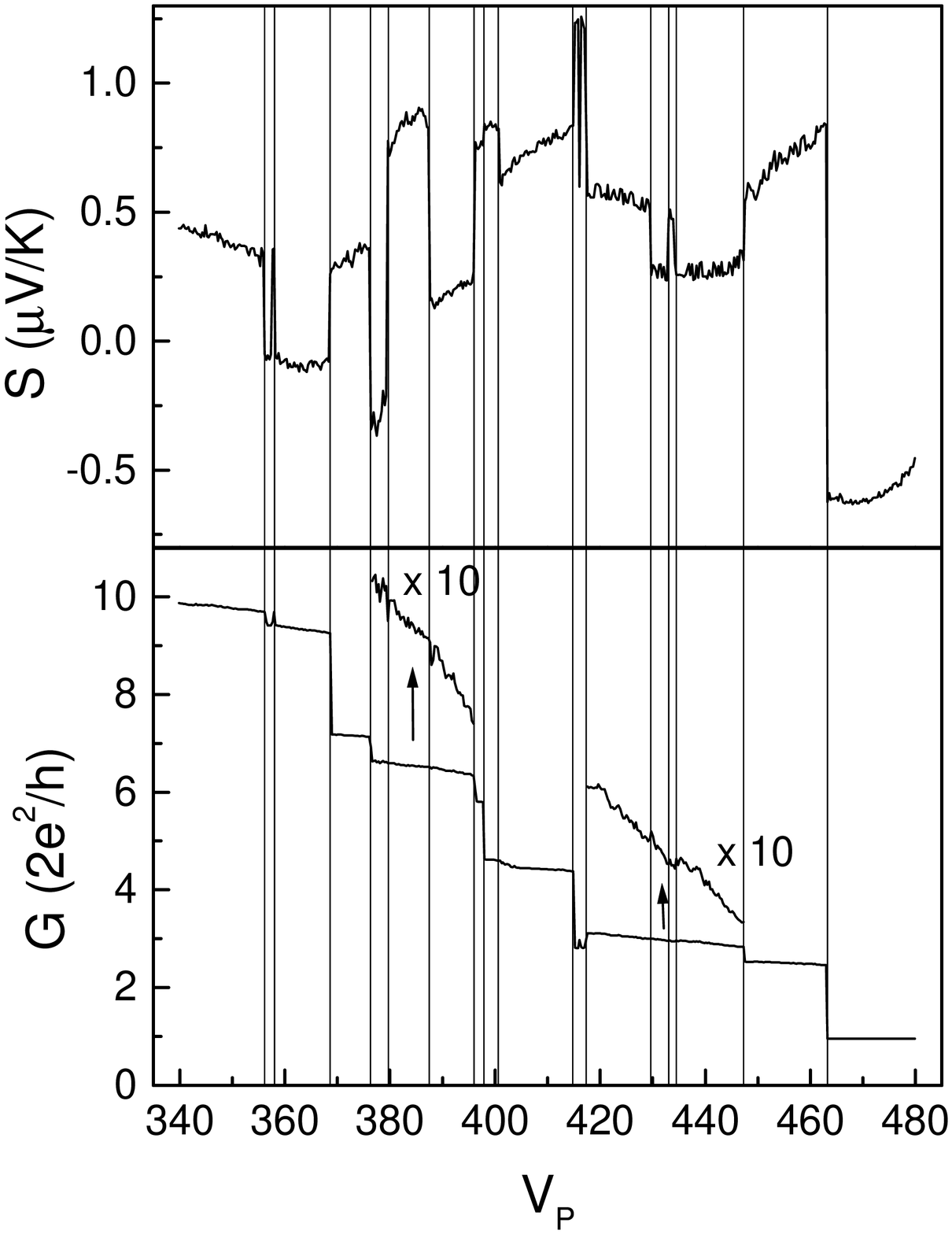,height=10cm,angle=0}
\end{center}
\caption{Typical conductance $G$ and thermopower $S$ versus piezovoltage 
$V_{P}$. The vertical gray lines indicate the corresponding steps in the 
conductance and thermopower. For two plateaus the conductance scale has 
been expanded 10 times and offset in order to show the small anomalies 
in $G$. }
\label{fig:fig2} 
\end{figure}

The primary limitation of the sample design is the thickness of the glass plates 
and thus the thermal insulation of the gold sample wire from the phosphor bronze 
substrate. The thickness is a trade-off between stability and thermal insulation. As a result of thermal currents flowing to the substrate, a thermal gradient is established in the sample wire between the thermometer and the contact. The measured temperature is hence not the actual temperature of the ``hot'' side of the constriction. We calibrate this thermal gradient by measuring the thermopower for large contacts, with resistances in the range 1 -- 10\,$\Omega$, as a function of heating power. First, we stress that the thermal 
resistance of the contact is orders of magnitude larger than that of the wire on 
either side, therefore the temperature difference over the contact can be taken 
to be independent of the contact size. This is corroborated by experiment, which 
shows that the mean value of the thermopower over the contact as a function of 
contact diameter in the range 0.1 -- 100\,$\Omega$ remains constant within an 
accuracy of 1\%. We take advantage of the fact that for conventional point 
contacts the phonon drag contribution to the thermopower becomes negligible 
\cite{shkly}. Since the contact is part of a uniform gold loop, the measured 
themopower corresponds to the phonon drag contribution of the leads only. The 
side that is not heated remains equal to the substrate temperature and we 
assume that the actual temperature difference over the contact is a fixed 
fraction, $\alpha$, of the measured temperature difference. We then determine 
this fraction by comparing the measured large-contact thermopower as a function 
of temperature with literature values for the bulk thermopower of pure gold 
\cite{pearson,guenault}, which is nearly linear between 10 and 25 K, with a 
slope of $\sim$0.05 $\mu$V/K$^2$. For the two gold samples discussed below, the 
model provides a good description when the fraction $\alpha$ is taken as 0.4 and 
0.5, respectively.  We estimate an error of about 20\% for the temperature 
difference obtained. 
\begin{figure}[!t]
\begin{center}
\leavevmode
\epsfig{figure=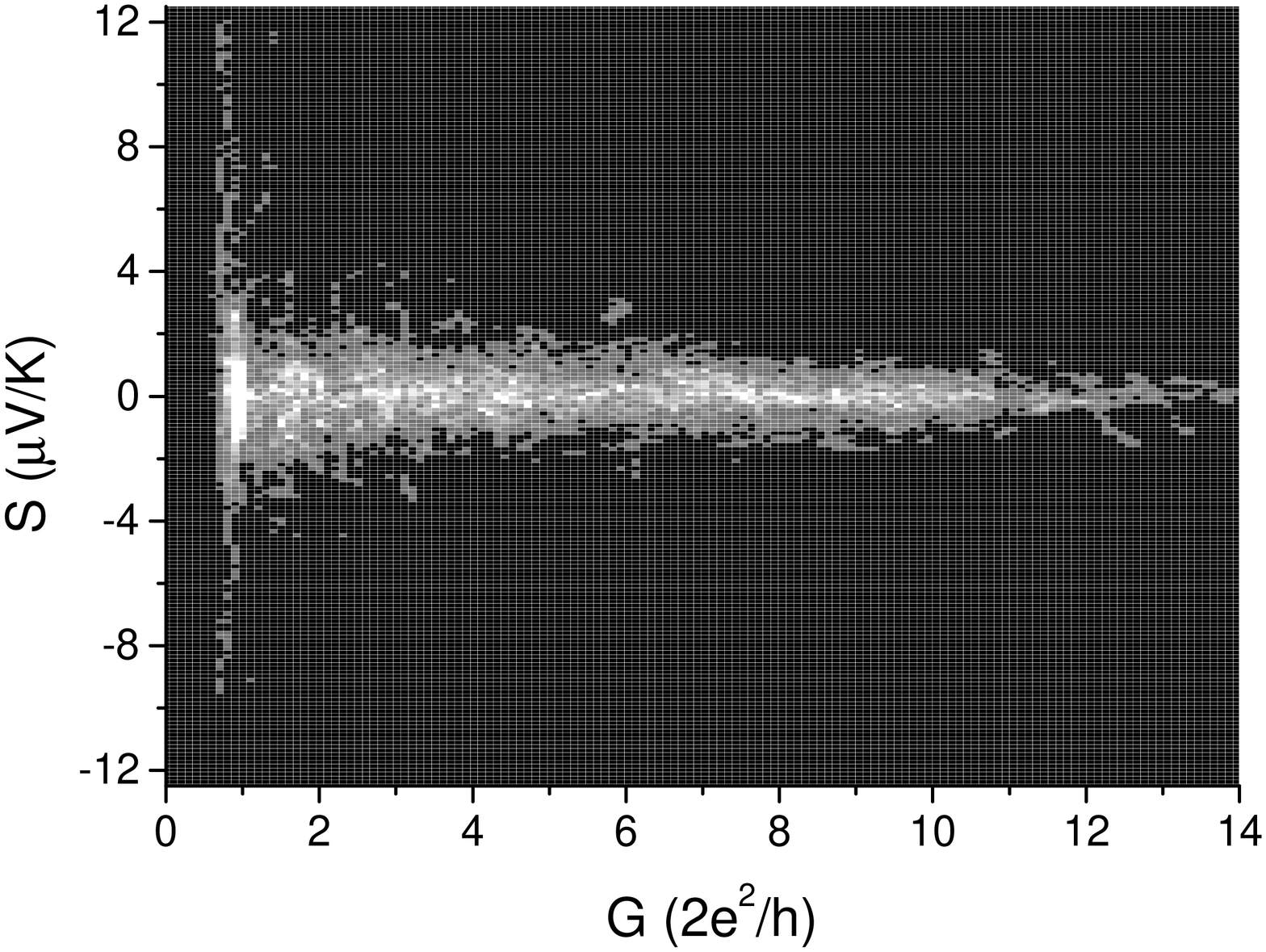,height=10cm,angle=270}
\end{center}
\caption{Density plot of thermopower against conductance: The 
thermopower axis is divided into 0.125\,$\mu $V/K, and the conductance 
into 0.1$G_{0}$ sections, creating 0.125\,$\mu $V/K by 0.1$G_{0}$ bins. 
The occurrence of a particular combination of conductance and $S$ is 
added to its corresponding bin, and the sum of 220 curves for the two 
samples is plotted in gray scale above. Black represents no data points 
and white more than 100. We note that above 10 $G_{0}$ fewer data have 
been taken.}
\label{fig:fig3}
\end{figure}

To have a reasonable signal level we need to apply a temperature difference of 
several kelvins.  In the case of sample 1 the temperature difference $\Delta 
\theta$ = 4\,K, and the average of the temperatures on both sides of the contact 
is $\theta_{av}$\,=\,11.5\,K. For sample 2, $\Delta \theta$\,=\,6\,K and 
$\theta_{av}$\,=\,12\,K. As it is not obvious that this $\Delta\theta$ 
can be regarded as a small (linear) perturbation, we will take the full 
dependence of $S$ on $\Delta\theta$ into account in the analysis below. 
For all the results presented below, the bulk thermopower of the leads 
has been subtracted. 

While breaking the contact by increasing the piezovoltage, the usual 
plateaus in the conductance are observed \cite{1atoms,rubio,muller,goldsteps}. 
When heating one side of the contact we observe steps in the thermopower which 
occur simultaneously with conductance jumps from one plateau to the next. 
Each measured curve produces a different conductance and thermopower 
trace and a typical example is shown in Fig.\,\ref{fig:fig2}. Even tiny 
jumps or changes of slope of the conductance can be accompanied by large steps 
in the thermopower. On a conductance plateau, even though the conductance hardly 
changes, smooth variations in the thermopower are usually observed. Note that 
the thermopower of the contact can have both a positive or negative sign. When 
we do not heat, or heat both sides of the contact to the same temperature, we 
observe no thermopower voltage within the noise level of 300\,nV peak to peak. 

In order to obtain statistical information about a possible correlation 
between the measured thermopower and conductance values, a density plot was 
constructed from the combined data of the 72 and 148 individual curves from 
sample 1 and 2, respectively (Fig.\,\ref{fig:fig3}). The conductance axis was 
divided into 10 partitions per $G_{0}$ and the thermopower axis in partitions of 
0.125\,$\mu$V/K. Then, the number of data points falling in each range 
of conductance and thermopower was counted and the results are represented 
in gray scale in Fig.\,\ref{fig:fig3}. In this figure we observe an increase in 
the spread of the thermopower values with decreasing contact size, with 
both positive and negative sign. 

A conductance histogram of the 220 curves is, within the statistical accuracy, 
in agreement with other gold histograms taken at low temperatures (e.g. Ref.
\cite{sirvent}). Although the data are not presented here, similar results for 
the thermopower have also been observed in silver and copper samples, albeit for 
a more limited number of curves.

The thermopower has a random value and sign, and seems to be much more sensitive 
to small changes in the atomic geometry of the contact than the conductance. 
This is not expected from the simple adiabatic models for point contacts, which 
only predict a positive sign \cite{thermtheory,bogachek}. We propose an 
interpretation of this behavior in terms of coherent backscattering of the 
electrons near the contact: As a result of the interference of waves with 
different path length, the transmission of the contact will show fluctuations as 
a function of energy \cite{fluct,ball}. Each atomic rearrangement at a conductance 
step will alter the interference paths of the backscattered electrons by a 
significant fraction of $\lambda _{F}$, and hence change the energy dependence 
of the transmission in an unpredictable way causing each step in the conductance 
to result in an unpredictable jump in the thermopower.  Along a plateau, the 
contact gradually changes position with respect to the scattering centers nearby 
and a gradual change in the interference pattern occurs. 

We now derive an expression for the thermopower based on these concepts. The 
thermopower in quantum point contacts can be written as $S=-L/G$, with 
\cite{thermtheory,bogachek}:
\[
\frac{L}{G}=\frac{\frac{2e}{h}\int_{0}^{\infty }(\text{Tr}\ {\bf t\, t}^{\dagger
})\ [f(\theta +\Delta \theta ,E)-f(\theta ,E)]dE}
{\frac{2e^{2}}{h%
}\int_{0}^{\infty } (\text{Tr}\ {\bf t\, t}^{\dagger})\frac{\partial f}{\partial 
E}dE},
\]
\noindent where $(\text{Tr}\ {\bf t\, t}^{\dagger})$ is the sum of the 
transmission probabilities, $f(\theta,E)$ is the temperature- and 
energy-dependent Fermi function, $h$ is Plank's constant, and $e=|e|$ is the electron charge. The thermopower is characterized by the energy dependence of the transmission 
probabilities.  We have approached the problem using the same method as presented in 
Ref.\,\cite{me} for the derivation of the conductance fluctuations in atomic 
size contacts. The point contact is taken as a ballistic central constriction 
with diffusive regions on both sides. The ballistic section (using the 
Landauer-B\"uttiker formalism) is characterized by $N$ conductance modes, each 
with a transmission probability $T_{n}$ and a contribution $T_{n}G_{0}$  to 
the conductance. After transmission through the contact, within the 
dephasing time $\tau_\phi$, the electron scatters elastically in the diffusive region and has a finite probability amplitude, $a$, to return to the contact. When the diameter of the contact is small compared to the elastic scattering length $l_{e}$, the return probability, $|a|^2$, is small and we need only consider lowest-order processes. To lowest order in $a$ we can write the transmission of the three sections combined as
\begin{eqnarray}
\text{Tr}\ {\bf t\, t}^{\dagger} = \sum T_n [1 + 2 \text{Re}\ (r_n a_{l,n} + 
r_n^\prime a_{r,n})]. \label{tr}
\end{eqnarray}
\noindent 
Here, $r_n$ and $r_n^\prime$ are the reflection coefficients in the 
transfer matrix describing the central ballistic section of the contact when 
coming from the left and right, respectively, with $|r_n|^2=|r_n^\prime|^2=1-T_n$. $a_{l,n}$ and $a_{r,n}$ are the return amplitudes for mode $n$ from the left and right diffusive regions, respectively.
The latter are sums over all possible paths of length $l$, containing 
phase factors $e^{i(E-E_{F})l/v_{F}\hbar}$. The second term in 
Eq.\,(\ref{tr}) describes the interference of the directly transmitted wave with 
the fraction that, after transmission, is first backscattered to the contact and 
subsequently {\it reflected} at the contact. Assuming the dominant energy 
dependence is in the phase factors, the integration over $E$ in the expression 
for $L$ can be performed. We consider the square of $L$, averaged over an ensemble 
of scattering configurations,
\begin{eqnarray} \label{L2}
\langle L^2 \rangle = \left( {2 e k_B \over \hbar}{\theta \over 
\Delta\theta} \right)^2 
\sum_n T_n^2(1-T_n) 
\int_{0}^{\infty } \langle |a(\tau)|^2\rangle 
\left( {1 \over \sinh(z)} -{1+\Delta\theta/\theta \over 
\sinh(z(1+\Delta\theta/\theta))} \right)^2 d\tau, 
\end{eqnarray}
\noindent
with $z=\pi k_B \theta \tau/\hbar$ and $\tau=l/v_F$. Here we have assumed that $a_{l,n}$ and $a_{r,n}$ are uncorrelated and have the same average return probability $\langle |a|^2 \rangle$, independent of the mode number $n$.

For $\langle |a(\tau)|^2\rangle$ we substitute the semiclassical probability to 
return to the contact into any of the $N$ modes after a time $\tau$, 
$\langle |a(\tau)|^2\rangle = v_F/[\sqrt{12\pi} k_F^2 (D\tau)^{3/2}]$, with 
$D=v_F l_e/3$ the diffusion constant. The integral in Eq.\,(\ref{L2}) can be 
performed numerically. It only weakly depends on the ratio $\Delta\theta/\theta$. For the standard deviation of the thermopower $\sigma_S=\sqrt{\langle S^{2}\rangle -\langle S\rangle ^{2}}=\sqrt{\langle L^{2}\rangle /\langle G^{2}\rangle }$ this finally results in
\begin{eqnarray}
\sigma_S= \frac{c k_{B}}{e\ k_{F} l_{e}\sqrt{1-\cos\gamma}}
\left( \frac{k_{B}\theta }{\hbar v_{F}/l _{e}}\right)^{1/4}
\frac{\sqrt{\sum_{n=1}^{N}T_{n}^{2}(1-T_{n})}}{\sum_{n=1}^{N}T_{n}}.  
\label{S}
\end{eqnarray}
\noindent
Here, $c$ is a numerical constant which equals 5.658 in the limit 
$\Delta\theta/\theta\rightarrow 0$, and increases by about 5\% for 
$\Delta\theta/\theta=0.5$. We have also introduced a factor ($1-
\cos\gamma$) to account for the finite geometrical opening angle of the contact, where the limit $\gamma$\,=\,90$^{o}$ corresponds to an opening in an infinitely thin insulating layer between two metallic half spaces. Note that $\sigma_S$ in 
Eq.\,(\ref{S}) is equal to zero when all $T_{n}$ are equal to either 0 or 1. 

\begin{figure}[!t]
\begin{center}
\leavevmode
\epsfig{figure=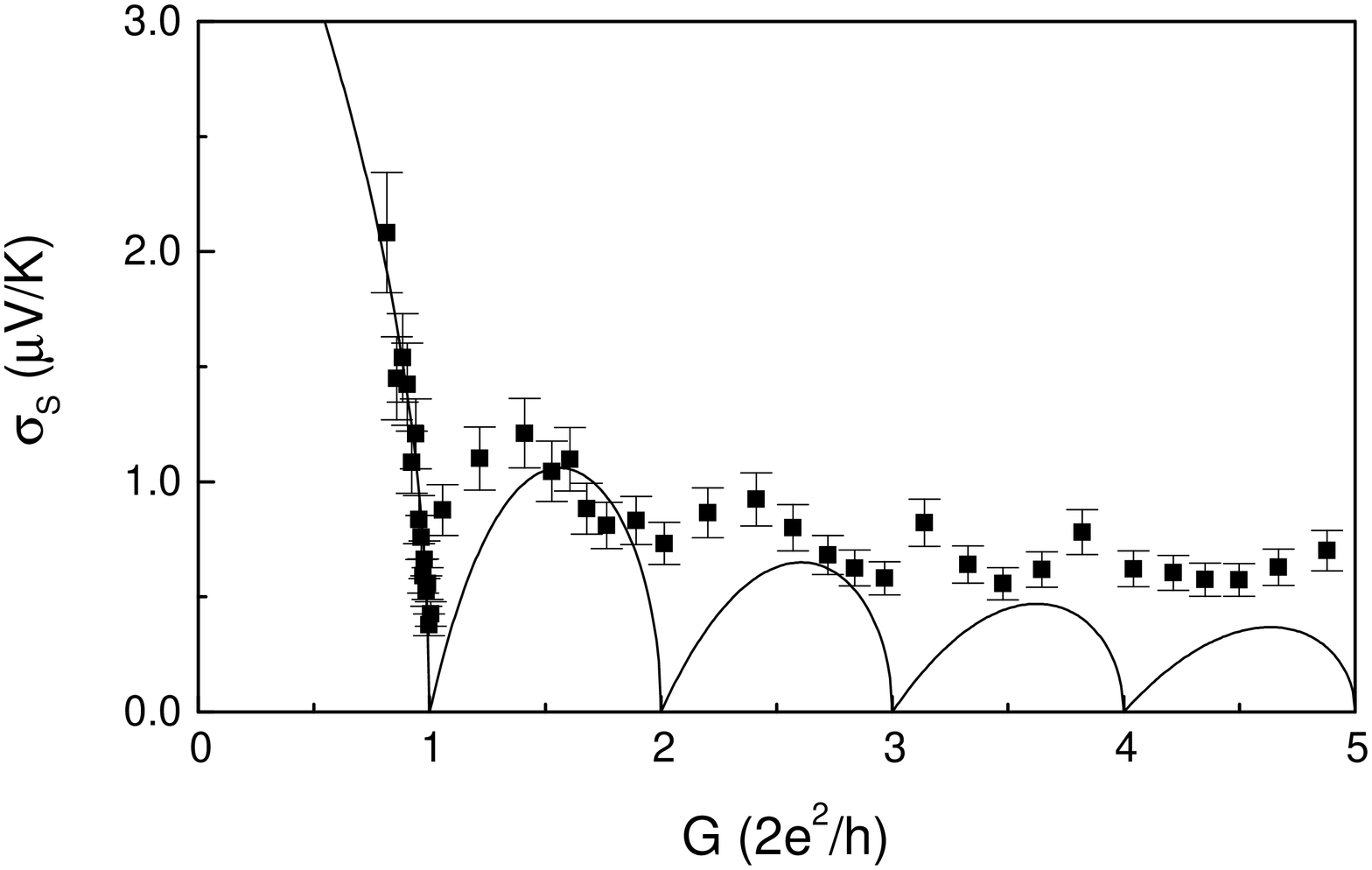,height=10cm,angle=270}
\end{center}
\caption{Standard deviation of the thermopower against conductance. 
Solid squares represent the measured data and the solid line is the 
theoretical curve assuming the conductance modes open one by one.}
\label{fig:fig4} 
\end{figure}

In Fig.\,\ref{fig:fig4} we plot  the standard deviation of the 
thermopower, determined from the experimental data by sorting all data points as a function of $G$ from the combined set of the 220 curves and averaging over 1000 consecutive data points. We compare these data to the theoretical curve calculated from the above expression for the case where the modes contributing to the conductance open one by one. That is, for all conductances with $N$ modes contributing to the conductance, only one (i.e., $T_{N}$) differs from 1 and $\sigma_S \propto T_{N}\sqrt{1-T_{N}}/(N-1+T_{N})$.  
The observed deep minimum at $G = G_{0}$ suggests that this conductance 
is dominantly carried by a single mode. This is in agreement with 
measurements of the shot noise on atomic size gold contacts \cite{helko}, and the observed suppression of conductance fluctuations in Ref.\,\cite{me}. For $G > G_{0}$, the limited statistics in combination with the property that the effect in the thermopower scales inversely with conductance prevent the definite 
identification of minima near quantized values. 

From the amplitude of the curve we obtain an estimate for the elastic mean free 
path of $l_e= 5\pm1$\,nm, using  reasonable values for the opening 
angle of the contact of 35$^{o}$--50$^{o}$ \cite{agrait2}. All data points should be on or above the full curve in Fig.\,\ref{fig:fig4}, since contributions by more conductance modes can only increase the variance in the thermopower. Therefore, $l_e$ cannot be much smaller than 4\,nm. For a much larger value of $l_e$ many modes would have to contribute to the conductance, in which case we would not expect to find any minima at quantized values.  

Apart from the thermopower effects described here, for a quantum point contact positive peaks in the thermopower, centered at conductance values $(n+{1\over 2})G_0$, $n$\,=\,0,1,2,..., were predicted due to the structure of the electron density of states \cite{thermtheory,bogachek}. This effect has indeed been observed in two-dimensional electron gas devices \cite{molenkamp}, but is much smaller than the fluctuations we observe here, and therefore cannot be resolved in the mean value $\langle S\rangle$ for our metallic point contacts.

The mechanism we present to explain the thermopower is the same as the 
one proposed for the voltage dependence of the conductance \cite{me}. 
Indeed, when we plot $\sigma_S$ and $\sqrt{\langle(\partial G/\partial V)^{2}\rangle}/G$ for 
gold, the data show very similar behavior. The energy scales with which both 
measurements have been performed are so different (6 K temperature 
difference versus 20 mV amplitude) that comparison between the parameters 
obtained by both methods is a test for the validity of the theoretical derivation. The mean free path value obtained from the conductance fluctuations is 5\,nm, in good agreement with the estimate obtained here. Hence, the fact that both works are not only in good qualitative but also in good quantitative agreement is strong support for the model used.  

\vspace{0.5cm}

This work was part of the research program of the \nobreak{``Stichting 
FOM''}, which was financially supported by NWO. We have greatly profited from 
many discussions with C. Urbina, D. Esteve, M.H. Devoret, and we thank 
L.J. de Jongh for his stimulating support.


\end{document}